\documentclass[numbers,sort&compress]{IntechOpen-Book}%%OPTIONS are numbers, authoryear, sort&compress,sectref

\graphicspath{{Artworks/}}

\usepackage{natbib}
\usepackage{fancyhdr}            % Permits header customization. See header section below.
\fancypagestyle{plain}{
    \lhead{}
    \fancyhead[R]{\thepage}
    \fancyhead[L]{}
    
    \fancyfoot{}
}
\begin{document}

\Mainmatter

\begin{frontmatter}

\chapter{Integrating MLSecOps in the Biotechnology Industry 5.0}
\title{Integrating MLSecOps in the Biotechnology Industry 5.0}
% Enter author names EXACTLY as you would like them to appear in the final manuscript; only use pattern: first name, last name. Do NOT use pattern: last name, first name. Patronyms fall into the first name category.
\author{Naseela Pervez$^{1}$}
\author{Alexander J. Titus$^{1, 2, 3}$}
% \author{Naseela Pervez$^{1}$ and Alexander J. Titus $^{1, 2, 3}$}

\makechaptertitle

\chaptermark{Integrating MLSecOps in the Biotechnology Industry 5.0}

\begin{abstract} % abstract = max 200 words, unstructured format, single paragraph
Biotechnology Industry 5.0 is advancing with the integration of cutting-edge technologies like Machine Learning (ML), the Internet Of Things (IoT), and cloud computing. It is no surprise that an industry that utilizes data from customers and can alter their lives is a target of a variety of attacks. This chapter provides a perspective of how Machine Learning Security Operations (MLSecOps) can help secure the biotechnology Industry 5.0. The chapter provides an analysis of the threats in the biotechnology Industry 5.0 and how ML algorithms can help secure with industry best practices. This chapter explores the scope of MLSecOps in the biotechnology Industry 5.0, highlighting how crucial it is to comply with current regulatory frameworks. With biotechnology Industry 5.0 developing innovative solutions in healthcare, supply chain management, biomanufacturing, pharmaceuticals sectors, and more, the chapter also discusses the MLSecOps best practices that industry and enterprises should follow while also considering ethical responsibilities. Overall, the chapter provides a discussion of how to integrate MLSecOps into the design, deployment, and regulation of the processes in biotechnology Industry 5.0.
\end{abstract}

\begin{keywords} % use a minimum of 5 kwrds, separate them with a comma
Artificial Intelligence, Machine Learning, Internet of Things, Security Operations, Biotechnology
\end{keywords}

\end{frontmatter}

\section{Introduction}
\label{intro}% first section MUST be titled Introduction, and feature introductory text; do NOT change this title
Industry 5.0 has revolutionized the way humans interact with various industrial sectors \cite{Adel2022-xx}, from human-machine collaboration to creating more human-centric experiences. With a growing emphasis on sustainability, ethical responsibilities, and involvement of social sciences in industry \cite{ethicsIndustry5.0}, there has been a growing awareness in big data, manufacturing, Internet of Things (IoT), digitalization, and production sectors. Industry 5.0 has witnessed the launch of humanistic technologies and products like human-approved high automation self-driven cars that leverage artificial intelligence (AI), which have been rolled out in various cities \cite{Lee2022-td}, IoT-based contact-less shopping experiences that seemed like need-of-the-hour during the pandemic \cite{Faang_2021}, and remote healthcare systems \cite{Zamzami2022-uf}.

The biotechnology sector is as relevant to Industry 5.0 as microprocessors are to computers. With every industrial revolution, the biomanufacturing sector has undergone constructive development. Industry 5.0 is expected to revolutionize the adaptation of robotic automation in manufacturing industries to provide a more humanistic experience while achieving high yields on production \cite{SELVAM2023108237}. As the biotechnology industry supports critical sectors like food, agriculture, healthcare, pharmaceuticals, and energy, companies must stay current with industry standards. Maintaining integrity, security, proper regulations, and ethical policies remain paramount to ensure responsible development and use of biotechnologies across industries \cite{Trump2022-hg}. Massive amounts of data in various biotechnology sectors (e.g. bioengineering, biodefense, forensics) have resulted in the deployment of technological solutions - Big Data, IoT, and AI in biotechnology. AI in biotechnology ranges from the use of machine learning (ML) models for predictive analytics \cite{Lee2006-fu} to employing generative AI techniques for accelerating discoveries with high computational power \cite{Caudai2021-kn}. AI, combined with IoT and Big Data, has resulted in an increasing presence of biotechnology data in cyberspace. To protect the data, we need to ensure that the AI models employed are secured. 

% \transition{To cybersecurity and machine learning in biotech}
% \naseela{The role of Machine Learning Security Operations (MLSecOps) in enhancing cybersecurity.}

Cloud technologies, which are centralizing data, have increased the threat of attacks \cite{Soveizi2023-vv}. There have been surveys and studies that have discussed the vulnerabilities of traditional cybersecurity practices \cite{Jang-Jaccard2014-mw, Kimani2019-wl}, and ML techniques have been created to enable more secure cyberspace. These tools are being developed to combat the growing threats to previously established cybersecurity techniques. Industry 5.0 has introduced state-of-the-art (SOTA) human-centric industry standards like employing robotic automation in manufacturing and production sectors, using IoT and AI-enabled healthcare systems, and employing generative AI in research and development. This has led to the need for developing better security measures including malware detection, network security, and intrusion detection; all crucial for the current industrial revolution. Machine Learning Security Operations (MLSecOps) integrates ML techniques to combat threats in the technology and digital space and highlights the cybersecurity vulnerabilities of deployed ML models. It is a promising approach to enhance cybersecurity \cite{Mohamed2023-wd} and promote a more secure, less vulnerable cyberspace for various industrial sectors. Machine learning has a proven ability to forecast future risks and threats, recognize threat patterns, and detect anomalies \cite{Apruzzese2022-ls}. These capabilities have led to the development of several ML-based cybersecurity pipelines \cite{Chachra2019-ab, Yamin2020-hi} in various industrial sectors.   

% \naseela{Objective of the chapter: To inform decision makers on the intricacies of MLSecOps in the context of Biotechnology Industry 5.0}

The presence of biotechnology data in cyberspace and the digitization of labs and biomanufacturing equipment have increased the industry's threat of attacks. The biotechnology sector must protect itself from cyber threats and cyber-attacks while also adopting the revolutionized approaches of Industry 5.0. Adopting MLSecOps to develop modern and effective cybersecurity measures to make the biotechnology industry less vulnerable and less prone to attacks will pave the way for digitalized biotechnology sectors (e.g. bioengineering, biosynthesis, biomanufacturing) In this chapter, we aim to provide an overview of existing vulnerabilities in the biotechnology Industry 5.0, integrating MLSecOps to secure the biotechnology Industry 5.0, and enabling decision makers to understand how ML techniques can be employed to secure the threat ecosystem of the biotechnology Industry 5.0, and how cybersecurity best practices can protect the rapidly expanding number of ML tools in production. Our focus is to provide an analysis and discussion of MLSecOps in the biotechnology Industry 5.0.

\section{Understanding the Biotechnology Industry 5.0}\label{biotech5.0}
% \naseela{Definition and Evolution: From traditional biotechnology to Industry 5.0.}
Biotechnology Industry 5.0 aims to employ advanced AI techniques to create more personalized products. Involving humans more in industrial processes while harnessing the power of advanced technology enables the industry to benefit from the best of human intelligence as well as advanced computation. Digitalization was introduced in industries during the third industrial revolution in the 1950s. This era was a catalyst for the biotechnology industry with the discovery of recombinant DNA which resulted in the development and production of new proteins, antibiotics, and modern-day synthetic biology \cite{Cohen1973-ue}. This was a significant step for the drug discovery sector in particular. The biotechnology industry has continued to digitize ever since. The commercialization of the first biotechnology drug, human insulin, which was developed using genetically engineered bacteria \cite{Johnson1983-oz} transformed the biotechnology industry from a niche industry into an economic engine. The intersection of biotechnology with modern technological discoveries like IoT, AI, and automation, and biotech sectors such as synthetic biology, genomics, and bioinformatics have accelerated research and development, manufacturing, and production industries across major economic sectors. With the biotechnology Industry 5.0, there is the potential for customization and personalization of biotechnology products and services (i.e. personalized medicine) powered by AI/ML, robotics, and IoT. This will not only improve the quality of human life in various aspects but it will also have a major impact on corporate growth.

% \naseela{Key Technologies: AI, ML, IoT, and their applications in biotech.
% }

% \naseela{Potential Impacts: On healthcare, agriculture, environmental sustainability, and bioeconomy.}

The role of emerging technology in the biotechnology industry is an area of interest for the decision-makers of numerous sectors \cite{biotechBioeconomy} and governments. In this section, we will highlight the key technologies and their impact on the biotechnology industry. We have briefly mentioned in section \ref{intro} the amount of data present in the biotechnology industry and how that has promoted the integration of the latest computational tools into the industry. Every sector of the biotechnology industry has an increasingly large amount of data that is used to draw meaningful insights, or that provides crucial insights for decision-making. This data needs to be stored and accessed by multiple organizations, either working on similar products or independent of each other. One of the advancements of Industry 4.0, was the centralization of data using \textbf{cloud computing} technologies. Cloud computing has enabled industries not only to store huge amounts of data cost-effectively but the data can also be accessed remotely by people in an organization. With the development of cloud technologies, data like genomics, biomarkers, and clinical data could be stored easily at low storage and maintenance costs \cite{Koppad2021-ex}. Data is considered a valuable asset to an organization, and companies have been collecting data for decades. With the higher computational power provided by cloud computing to study and analyze the data, the data is worth billions of dollars in today's world \cite{Haggin2021-ge}. The data in the biotechnology industry has been used for a plethora of sectors - genetic engineering, personalized medications, regulatory compliance, sustainability, and identifying patterns in genomic data, among others.

\textbf{Artificial Intelligence} (AI) is used to develop computer systems that can process and recognize patterns in data, draw statistical inferences, and interact with the physical world through mechanisms such as IoT devices \cite{Xu2021-gh}. In biotechnology, AI is being employed to support decision-making in the development of smart and intelligent systems. \textbf{Machine Learning} (ML) is a sub-discipline of AI that employs statistical methods for drawing inferences from data. ML models have been used to detect COVID-19 virus from MRI's \cite{Jia2022-ir}, optimize manufacturing processes \cite{Rajesh2022-uw}, analyze crop growth patterns in agriculture \cite{Sahu2023-aw}, and gain insights from gene data \cite{Ahemad2022-kg, Schwarzer2021-cy}. Robotic automation of biotech labs has resulted in a reduction in human errors, improving safety, enabling faster clinical translations, and increasing the rates of experimentation \cite{Holland2020-ku}. ML and deep learning (DL) techniques have furthered the development of biotech industries since Industry 4.0. Recent developments in AI, like generative AI, explainable AI, and large language and vision models (LLMs) have opened up applications of AI/ML in the biotechnology industry. 

Environmental factors play a crucial role in the biotechnology sector. For example, precipitation and sunlight play a significant role in the agriculture sector in determining crop yield and temperature is a major factor for drug production. Keeping these environmental factors under control requires human intervention. \textbf{Internet of Things} (IoT) was introduced in the biotech industry to reduce human labor and to control environmental factors. There is a study that has proposed an end-to-end web application-based framework that integrates embedded Arduino systems with Abiotic sensors to monitor, measure, and analyze environmental factors like temperature, humidity, luminance, audio/noise, and CO levels and precision manufacturing is a rapidly growing market \cite{Vidakis2017-vz}. 

Additionally, IoT devices are integrated into healthcare systems for medical devices and remote patient monitoring \cite{Yew2020-yg} which has promoted patient care from the comfort of their homes. One of the recent breakthrough developments in IoT devices is the wearable device.  Wearable technology and digitalization of healthcare systems can promote patient engagement and involvement in their healthcare plan \cite{Bove2019-gq}, and these devices are used for daily monitoring of physical factors like heartbeat, oxygen levels, and human activity. 

Industrially, the use of IoT devices in various sectors of the biotechnology industry has enabled the expansion of biomanufacturing capacity, giving a boost to the bioeconomy \cite{Wei2022-dq}.

This section has briefly described the effect various technologies have on various biotechnology industries. The biotechnology Industry 5.0 focuses on developing a more humanistic approach to utilizing these technologies. IoT devices in combination with AI can unlock powerful capabilities like personalization, automation, context awareness, enhanced forecasting, and predictive maintenance. Data being at the center of all this creates targets and vulnerabilities in the biotechnology Industry 5.0. The subsequent sections will discuss the threat landscapes, safety regulations, and enabling AI/ML techniques to ensure a secure biotechnology framework.

\section{The Role of MLSecOps in the Biotechnology Industry 5.0}\label{sec3}

The biotechnology Industry 5.0 combines digital technology with biotechnology. Industry 5.0 has made breakthrough solutions to complex solutions - automatic therapeutic systems \cite{Tomassini2023-kr}, anti-aging research \cite{Duan2022-xg}, cancer treatment research \cite{Ayala-Orozco2023-ts}, lab-grown organs \cite{Li2022-bq}, among the most impactful ones. These complex experiments have led to data being at the center of the biotechnology research and development sector \cite{Tommasone2023-ib}. There is a persistent tension in biotechnology - threat and safety. Like any other industry, ML algorithms are being applied to develop efficient solutions. Section \ref{biotech5.0} provides an overview of how ML has been used to architect various design solutions to biotechnology problems. Section \ref{threat_landscape} will dive into threats in the biotechnology industry. It will explain where security breaches can occur and how they affect the implementation of digital technology in the biotechnology sector. This section is a discussion on the role of ML in security operations in the biotechnology Industry 5.0. This section provides two distinct views:
\begin{enumerate}
    \item Employing ML techniques to secure and protect the biotechnology industry against cyber threats.
    \item Securing the machine learning models against attacks employing cyber principles.
\end{enumerate}

Machine learning (ML) has been employed to enhance the performance of various traditional technologies. Autonomous vehicles are a great example of this. Although ML algorithms play a major role in product development, they have promising applications in other sectors like deployment and security. 

Genome data have been growing at an unprecedented pace \cite{Stephens2015-ye} and are used in various industries - forensic investigations, healthcare, and biotechnology being primary users. Preserving genome data is crucial and one of the most important tasks currently at hand in the biotechnology Industry 5.0. The current legal frameworks aim to protect genome data from malicious attacks but the data is still the target of existing threats \cite{Wan2022-ud}. Encrypting genome data while sharing and accessing it provides some protection against attacks, but encryption algorithms are prone to being attacked themselves by hackers to gain access to data \cite{Kuo2022-rp, Arshad2021-jc}. Genome data, if hacked can be manipulated and used for harmful activities. ResNetAct is a deep learning model based on the pre-trained ResNet model that has been fine-tuned to encrypt genome data \cite{Song2023-lf}. AI/ML algorithms have also been successful in detecting threats and ransomware detection \cite{Charmilisri2023-fg}. AI/ML algorithms have shown promising performance on the classification task of the ransomware samples \cite{Almousa2021-ke}, and deep learning architectures have been employed to detect Trojan attacks on encrypted DNA data. The supply chain is a also crucial sector of the biotechnology Industry 5.0 and directly affects every aspect of the bioeconomy \cite{Islam2022-tw}, from transportation to industrial machinery in labs. It even extends to automating fault detection and building resilient fault detection systems that keep highly sensitive research and manufacturing equipment safe. 

ML techniques have been integrated into fault detection pipelines in automobile supply chain frameworks. Statistical ML algorithms like linear discriminant analysis and Naive Bayes algorithm have shown promising results in classifying the condition of electric drive trains in supply chain \cite{Lahmiri2023-bu}. ML algorithms are also employed for monitoring and assessing bioprocesses in the biotechnology Industry 5.0. \cite{Mondal2023-tm}. Biotechnology Industry 5.0 has a network of software in various sectors \cite{Ben_Youssef2023-vb} and everything from running experiments to carrying out visualizations and maintaining lab inventory all have different software today. These software are prone to attacks and have vulnerabilities in code that make them less secure. Large language models (LLMs) have showcased promising results in identifying vulnerabilities in code \cite{Omar2023-wg} which can be employed in securing the software architecture. 

Healthcare and pharmaceuticals are major contributors to the bioeconomy. The application of ML to diagnostic procedures has been booming. However, developing these solutions while keeping data privacy and protection in mind is crucial. Federated machine learning algorithms have been used for developing secure medical imaging diagnostic solutions \cite{Kaissis2020-jd}. The bioinformatics sector of the biotechnology Industry 5.0 is also vulnerable to attacks, and biometric recognition systems are particularly impactful if breached, as these systems rely heavily on AI/ML algorithms. The biotechnology Industry 5.0 is emerging in developing robust authentication systems by training advanced ML models on real-time data (e.g developing an object recognition system on real-time humans rather than images of humans which might classify a human sticker on a road as a pedestrian \cite{Salunke2023-hn}) as well as using ML for continuous authentication of mobile devices based on behavioral biometrics \cite{Rocha2020-nu}.

Throughout the chapter, there have been examples showcasing the use of ML in combination with IoT devices in the biotechnology Industry 5.0. These models are used for an array of solutions in the biotechnology industry, including predictive analysis for crop yield \cite{Elbasi2023-la},  classification of genome data \cite{Monaco2021-vd}, enhance manufacturing of biotechnology products \cite{Hickerson2023-kl}, management of biotechnology labs \cite{Duong-Trung2023-es}, and more. Like other components of Industry 5.0, the ML models are also vulnerable. ML models are like intellectual property. If one gets access to data for gene therapies, emulating results or altering the therapy is a concern for genetic research and development \cite{Lux2017-ys}, and there have been studies carried out to identify the vast threat landscape in ML models \cite{Tidjon2022-to}. MLSecOps in the biotechnology Industry 5.0 not only uses ML techniques for security operations but also protects the AI and ML models that are used to design solutions or products. ML models, particularly modern AI applications, require huge computational power for training. Most of the training of these huge models is done on outsourced cloud platforms, and they are increasingly being subjected to malicious training \cite{Gu2017-ii} which can lead to drift in the expected results. The above-mentioned paper has carried out 2 experiments - 1. handwritten digits and 2. street sign detector, which reflect that the model performs well on user data but badly on some inputs that might be attacker chosen e.g. a situation where red light can be classified as green. One of the breakthroughs in Industry 5.0 has been the use of language models in all sectors. The large language models (LLMs) are used for automating production, customer service, as well as automating instruction manuals in manufacturing industries. However, these models are prone to attacks. Data theft is one of the major concerns and is one of the reasons for skepticism to applying ML models in the biotechnology Industry 5.0 \cite{Pavaloaia2023-ii}.

Frameworks have been developed to identify and mitigate backdoor triggers in language models. One of the ways through which the model training can be protected against malicious training samples is to train a second model to recognize if the triggers are classified as safe targets \cite{Omar2023-ad}. Preventing corrupted samples from training subsets has shown impressive results in securing the language models and building the confidence to employ these secure language models in production lines of supply chain framework for various industries. Bias in machine learning models is also an active research area. These biases can be reflected in the results produced by ML systems. Integration of ML techniques in the healthcare industry has reflected biases in the electronic health records of the patients \cite{Gianfrancesco2018-ub}. Identifying and reducing these biases is crucial to make the ML systems trustworthy in biotechnology Industry 5.0. There have been solutions developed to tackle the problem of model bias in the healthcare sector of biotechnology Industry 5.0, where focusing on data handling for ML systems has been shown to reduce these biases in radiology data \cite{Rouzrokh2022-pa}. Another approach to identifying the biases and risks of ML models is to perform behavioral analysis of the model on outlier data points and keep track of the model performance for any abnormalities \cite{Cabrera2023-zn}. Analyzing model metrics and behaviors over time enables users to understand if there are any biases or alterations in models that affect the model performances (e.g. if a classifier's accuracy decreases over time or classifies input wrong, it indicates that the model has been altered). Monitoring frameworks in biotechnology Industry 5.0 have huge applications and requirements to catch any abnormal ML model behaviors early on. 

In this section, we have discussed the role of MLSecOps and its application in the biotechnology Industry 5.0. We have provided an overview of how ML systems can be used to protect biotechnology Industry 5.0 frameworks. This section also emphasizes the importance of securing the ML frameworks and infrastructures employed in biotechnology Industry 5.0. The remainder of the chapter focuses on identifying the threats in the biotechnology Industry 5.0 and discussing regulations and ethical responsibilities that should be fulfilled while designing infrastructure and solutions for various sectors in the biotechnology Industry 5.0. 

\section{Threat Landscape in Biotechnology Industry 5.0}\label{threat_landscape}
Biotechnology Industry 5.0 can lead to progressive research as well as product development in the biotech sectors. This is crucial for the bioeconomy. We have discussed the biotechnology sectors and the effect of the fifth industrial revolution on the biotechnology industry in section \ref{biotech5.0}. Although employing combinations of modern technologies like advanced IoT devices and ML models that can outperform human expertise has resulted in smart industries \cite{Herath2022-vj}, it has also exposed biotechnology data to threats. Section \ref{biotech5.0} has also introduced cloud technologies in biotechnology Industry 5.0. The presence of huge amounts of data on these centralized systems and the growing reliance on production ML models has increased the vulnerability to cyber-attacks. This section covers the biotechnology industry's cyber threats as well as sophisticated cyber attacks. We will provide references to case studies that are crucial in the cybersecurity space in biotechnology Industry 5.0. Recognizing threats specific to biotechnology Industry 5.0 will help stakeholders make informed decisions on securing the industry and introducing security measures in biotech sectors.

\subsection{Cyber Threats Specific to Biotech: Identifying unique vulnerabilities.}

There is variability in the data in biotechnology Industry 5.0, which requires unique and specific attention to these vulnerabilities. \textbf{Digital DNA} has wide applications in retail, healthcare and forensic departments \cite{noauthor_2020-he}. Encoding DNA efficiently into digital form opens the horizons to employing AI/ML algorithms on this data for various downstream tasks (DNA sequencing, DNA genotyping, and DNA matching ) \cite{DePristo2011-bv, Busia2018-xx}. Sensors and IoT devices have become a part of our livelihoods. Homes have light sensors for energy conservation \cite{Sithravel2024-xw}, cars have motion sensors to detect real-time vehicle motion \cite{Zhao2019-so} which can be employed in self-driving cars, and thermal and humidity sensors have been used to check for overheating in buildings \cite{Szagri2022-xx}. Similarly, the use of IoT devices in various biotechnology sectors like agriculture, healthcare industries, and biomanufacturing can create end-to-end cost-effective, and high-efficiency systems. Wireless Body Area Network - a network of IoT devices or sensors as \textbf{wearable devices} as well as \textbf{implants}, have revolutionized remote healthcare, making patient real-time monitoring systems proactive \cite{Pramanik2019-rs}. This can have promising results in developing healthcare systems for patients, such as those with dementia, that maintain a balance between the risks of symptoms like wandering while also assuring autonomy to the patients \cite{Robinson2007-up}. In addition to healthcare systems, there have been studies and reports in the agriculture and food domain carried out for multiple use cases \cite{Verdouw2019-fv}. The data from these implants and IoT devices are used to create dashboards for tracking and management purposes, ML architectures for drawing out various inferences or stored in the cloud to be accessed by stakeholders. This makes the data from IoT devices vulnerable and also poses a risk of breaching the privacy of the stakeholders involved. \textbf{Biometrics} like fingerprints and facial recognition systems are used in defense and law enforcement industries as well as other industries and unauthorized access to biometric data can aid bad actors, such as terrorist organizations, and create a risk to an entire nation \cite{Guo2021-mp}. Securing biometric data like facial features, iris patterns, speech data, etc is of national interest and there have been attempts to secure this data \cite{Shopon2021-iw}. Digital ID cards like driver's licenses as well as state IDs have made storing and producing these cards an easy task. Banking services have automated their identification system using AI-enabled technologies. This means an AI model will access your driver's license to check for authenticity. ID verification using AI has been viewed as a hazard to human privacy. A biometric data breach can lead to identity theft, deepfakes, and possible criminal activities. Biotechnology Industry 5.0 has to ensure that the data in various sectors of the industry is secure and protected against malicious attacks.

\subsection{Advanced Persistent Threats (APTs) and Biohacking: Addressing sophisticated cyber attacks.}

The COVID-19 pandemic witnessed the need for protecting the biotechnology Industry 5.0. Pandemics and epidemics pose a risk to a broad swath of society. \textbf{Advanced Persistent Threats (APTs)} can be described as malicious activities that are hard to uncover and remain undetected. In the biotechnology industry, one example of APTs is illegally manufacturing and selling some variants of approved drugs \cite{Mustazza2018-md}. The under-the-radar behavior of APTs allows them to remain undetected for longer amounts of time. In the past, there have been reports of APTs attacking various sectors of the biotechnology industry. Biotechnology Industry 5.0 being more than ever present in cyberspace makes it prone to APTs. Currently, various APTs have been reported to be active. One of them is APT41, suspected to have originated in China and targets biotechnology sectors like the pharmaceutical industry for financial gains \cite{fraiser_2021-un}. The healthcare industry has always been a threatened sector of the biotechnology industry. Case studies have been carried out that showcase recent developments to make the threats more evasive by Wekby (also known as APT18) - a famous APT in the healthcare sector \cite{aaron22}.

% \naseela{Case Studies: Real-world examples of cyber threats in biotech \textbf{I have included the related case-studies in the above sections}
% }

In this section, we have provided a picture of the biotechnology Industry 5.0 data that could be under attack and needs to be secured and protected. We have also given a brief on the progression of APTs and biohacking in the biotechnology Industry 5.0. Now that we have a deeper understanding of the biotechnology-specific threat landscape, we can focus on how developing and adopting practices that avoid these landscapes. The rest of the chapter will focus on how MLSecOps can be integrated into biotechnology Industry 5.0 to enhance privacy and security. We aim to discuss how MLSecOps can be employed for cybersecurity while ensuring that we maintain regulatory adherence and consider social and ethical responsibility.

\section{Regulatory Frameworks and Compliance}\label{sec5}
The biotechnology Industry 5.0 has millions of consumers throughout the world. It has an approximate value of \$2.9 trillion impact on the US economy and provides 2.1 million jobs in the United States \cite{Thacker2022-di}. The biotechnology sectors include agriculture, healthcare, pharmaceuticals, energy, and many sectors that have a direct impact on consumers. The products of the biotechnology industry must meet standards of quality, privacy, and security \cite{Esearch_undated-kc}. Regulatory frameworks are established to standardize and streamline production and manufacturing processes and prevent any security or privacy breaches. In the previous sections, we discussed how ML and AI are integrated into the security operations of the biotechnology Industry 5.0. We have discussed the current threats to the biotechnology Industry 5.0 in section \ref{threat_landscape}. Throughout the chapter, we have discussed how IoT in combination with ML and AI can help personalize the biotechnology Industry 5.0. However, enterprises are apprehensive about rolling out AI-based products to the public \cite{McKendrick2022-bo}. These products must achieve complete compliance with regulations. In this section, we provide an overview of the current regulatory frameworks, challenges in compliance frameworks, and standardization of regulatory and compliance frameworks globally, keeping biotechnology Industry 5.0 as a focus point.

\subsection{Current Regulatory Landscape: Overview of existing cybersecurity and biotech regulations.}
\label{cybersecuritty_regulations}
Artificial Intelligence integration and the growing digitization in the biotechnology Industry 5.0 have made the industry vulnerable to cyber attacks. To ensure the safety of the industry as well as to ensure that consumers' privacy stays intact, the biotechnology Industry 5.0 has to comply with 2 major groups of regulations  - \textbf{biotechnology regulations} and \textbf{cybersecurity regulations}. Here we discuss what regulations fall under these that the biotechnology Industry 5.0 has to comply with. Biotechnology regulatory frameworks include the sectors of the food industry, drug production, energy industry, etc. The National Academies of Sciences Engineering and Medicine \cite{National_Academies_of_Sciences_Engineering_and_Medicine2017-ou} has provided an overview of existing regulations in these sectors. The food and drug industries are very large consumer markets. They directly affect children, adults, and seniors equally. In the past, there have been mishaps related to the food industry. In 2015, there was a multi-state outbreak of high-risk infection Listeriosis in the United States that was directly linked to Blue Bell Creameries products \cite{noauthor_2018-fz}. There have been reports of 23 deaths among patients who are using a non-approved anti-gout drug \cite{noauthor_undated-si}. Developing and adapting the regulatory frameworks in the food and drug industry is crucial. The \textbf{Food and Drug Administration (FDA)} is a prominent regulatory body of the food and drug industry in biotechnology \cite{noauthor_2024-op}. FDA regulates and approves products like food, drugs, medical devices, vaccines, cosmetics, tobacco, etc. Biotechnology Industry 5.0 will see a boom in technologically advanced products, especially medical devices. For these devices to be approved for public use, they must comply with FDA regulations. FDA also regulates gene therapies which are crucial to the biotechnology Industry 5.0. 

While building products that comply with regulatory frameworks is essential, there are also regulatory frameworks that oversee the production process. Bioprocessing and biomanufacturing industries release substances into the environment. Global greenhouse gas emissions have led to climate change which affects terrestrial as well as aquatic organisms. Wastes from the manufacturing sector of the biotechnology Industry 5.0 include biological waste like pathogens, toxins from drug production, heavy metals, hazardous chemicals, and greenhouse gases. Industry 5.0 needs to ensure sustainability while developing high-efficiency products. The \textbf{Environmental Protected Agency (EPA)} ensures proper waste disposal, regulates the production of hazardous gases and chemicals, and regulates the environmental release of pathogens \cite{Us_Epa2013-uc}. The supply chain management of biotechnology Industry 5.0 needs to ensure that all the processes comply with EPA regulations. 

Agriculture is one of the largest sectors of the economy globally. The agriculture sector of biotechnology has prospered and developed into an economic asset for every nation. However, there are regulatory frameworks that the sector has to comply with to ensure safety and protection. \textbf{US Department of Agriculture} \cite{noauthor_undated-ff} overlooks the agriculture industry and regulates farming, food and nutrition, forestry, opioid production, and organic produce frameworks. The USDA also ensures that it complies with EPA regulations and maintains the climate conditions. 

Having discussed some of the major biotechnology regulatory frameworks, some cybersecurity frameworks are crucial to the biotechnology Industry 5.0. Medical devices and monitoring products have been growing more than ever. Remote patient monitoring is a prospective advancement in biotechnology Industry 5.0. However, these systems should be safe and secure for the patients. \textbf{Health Insurance Portability and Accountability Act (HIPAA)} \cite{noauthor_2022-yu} ensures that healthcare industry protects the patient information. HIPAA ensures that the healthcare products released into the market safeguard the patients' privacy. Business owners who step into the healthcare sector should ensure HIPAA compliance. Online payments and transactions have made our everyday lives easy. Any purchases can be made online without any hassle using your card. However, card information needs to be protected. Biotech labs and healthcare businesses make use of online transactions. However, the payment gateways must secure the card information. \textbf{Payment Card Industry Data Security Standard (PCI DSS)} \cite{noauthor_undated-de} ensures the protection of card information data while making online payments. All enterprises in the biotechnology Industry 5.0 need to comply with this standard to ensure consumer trust. The \textbf{National Institute of Standard and Technology (NIST) Cybersecurity Framework} \cite{Nist2023-wu} provides guidelines to enterprises, industries, and government agencies to minimize cybersecurity risk. The framework ensures a high level of protection against cyber attacks. Complying with NIST cybersecurity regulations will ensure that advanced technology can be employed in biotechnology Industry 5.0 with more confidence, fewer apprehensions, and fewer vulnerabilities. As we transition into Industry 5.0, the need for trustworthy, safe, and secure AI practices is needed more than ever. The US government has released executive orders that outline the emerging policies for the safe and secure use of AI \cite{The_White_House2023-gn}. The executive order has mentioned practices and policies that ensure the safe use of AI by developers as well as protect the public from AI-enabled attacks. This is truly the need of the hour as we see AI being used in all dimensions of industries and integrated into public use at a scale that hasn't been witnessed before. 

\subsection{Compliance Challenges: Adapting to fast-evolving biotech advancements.}

Although the above-mentioned regulatory bodies ensure security and safety, complying with these regulations has put many enterprises and industries in a tight spot. From product development to product release, complying with so many regulations requires a separate legal team, abundant documentation, and iterating through the process over and over again. Enterprises, especially small business owners, believe that following so many regulations is counter-productive and costs them money as well as labor \cite{Shapiro2020-tc}. The study further presented the views of how small business owners have a hard time making decisions to comply with regulations due to the absence of behavioral heuristics taken into account while designing these frameworks. The regulatory frameworks differ from country to country. And so does the expense and documentation of implementing these frameworks. Business owners in the United States have reported higher legal costs and delayed economic activity or profit to get a compliance clearance from the government as opposed to other economically established nations  \cite{Potter2002-er}. 

A key example is the discovery and development of the ground-breaking tool, CRISPR, for gene modification. CRISPR has potential applications in agriculture to cater to global food demand and human healthcare to develop treatment plans for genetic diseases and cancer \cite{Nidhi2021-vt}, among many other areas. CRISPR combined with AI technologies will transform medicine in the next decade, and while CRISPR was first published in 2012, the first evidence of CRISPR being fruitful in gene therapies was identified in under a decade \cite{noauthor_undated-fc}. However, CRISPR has only just been approved in December 2023 by the FDA for sickle cell disease \cite{noauthor_undated-fh}. There is a major opportunity for CRISPR, and other emerging biotechnologies, to have an impact as long as regulation modernizes alongside it.

Many technologies discussed in this chapter utilize IoT and AI sectors like machine learning, deep learning, and reinforcement learning. However, getting these products to production lines has been a struggle for biotech companies because they do not meet the current regulatory standards. Although there are compliance challenges currently in biotechnology Industry 5.0, there is no doubt that Industry 5.0 and its ground-breaking developments are here to stay. Transitioning from Industry 4.0 to Industry 5.0 relies on the decision making of policymakers to set standards for regulatory frameworks \cite{Ghobakhloo2023-bq}.

\subsection{International Standards: The role of global cooperation in standard setting.}
A common refrain notes that the \textit{world is a small place}. We can now travel from one side of the world to another in a matter of a few hours. Standardization has been implemented in measuring distances, heights, and weights for many centuries. Now that businesses are globally releasing their products, they have to make sure that their product meets the regulations of each country they appear in. This costs the business owners time, money, and labor. Introducing technology in all industries has challenged global standardization. Technology has had a global impact on standardization policies \cite{Holland2023-ed}, with nations largely developing their own standards. Standardization ensures interoperability, which means that a product developed in Italy can benefit the people of the United States, and vice versa, equally and unified standards would be beneficial in the biotechnology industry. Standardization maintains consistency in the industry while also ensuring that consumers are protected and there is transparency between the industry and consumers \cite{Holland2023-np}.

We now have a good understanding of how biotechnology Industry 5.0 differs from biotechnology Industry 4.0 and what are the potential threats in the biotechnology Industry 5.0. We have also discussed how ML can be implemented in the security operations of biotechnology Industry 5.0. This section gives an overview of the existing regulations and how Industry 5.0 has compliance challenges with these regulations. We have also briefly highlighted the importance of maintaining global standards. In the remainder of the chapter, we will present a perspective of how can MLSecOps be integrated into biotechnology Industry 5.0 while complying with the regulations, and the ethical and social responsibilities of Industry 5.0. We will also provide a perspective of what goes into decision-making for designing policies and regulations for Industry 5.0 focusing more on biotechnology Industry 5.0.

\section{MLSecOps Best Practices for Biotechnology Industry 5.0}\label{sec6}

Industry 5.0 is all about personalizing products and making industrial processes human-centric. In section \ref{biotech5.0} and section \ref{sec3}, we have discussed the role of machine learning in Industry 5.0 focusing on the biotechnology sector. The chapter has highlighted the threat landscape in biotechnology Industry 5.0 in section \ref{threat_landscape} which is one of the most crucial aspects of employing the latest technology. Having discussed the role of machine learning in MLSecOps and the current threat landscape in biotechnology Industry 5.0, as well as the major regulatory bodies in section \ref{sec5}, this section aims to discuss the MLSecOps practices that can be employed in biotechnology Industry 5.0. In this section, we aim to discuss how MLSecOps can be integrated into Industry 5.0 while also complying with current regulatory frameworks. In section \ref{sec3} we have discussed the role of ML in MLSecOps in two aspects - protecting the machine learning systems used in Industry 5.0 and employing machine learning to protect the Industry 5.0 products. In this section as well we will discuss the MLSecOps practices keeping the two aspects of machine learning in MLSecOps for biotechnology Industry 5.0. 

\subsection{Risk Assessment and Management}

Risk assessment and management is a major part of all sectors. Building products with risk analysis and management best practices ensures that the complexities in manufacturing, production, and usage have been considered before the products are commercialized \cite{Rodriguez-Espindola2022-xy}. Biotechnology Industry 5.0 includes sectors like informatics, pharmaceuticals, healthcare, and agriculture. These sectors all have crucial data and information that needs to be protected against threats. The products in Industry 5.0 employ machine learning techniques in various processes which makes Industry 5.0 more vulnerable \cite{Duarte2022-np}. Risk assessment and management ensure that these products protect the privacy of the enterprise, industries, as well as end users. The healthcare sector in biotechnology Industry 5.0 deals with high risks of phishing, cloud configuration errors as well as web application-based attacks, among others. With the increased use of modern technologies like IoT, cloud, and artificial intelligence, healthcare data is increasingly online. There have been studies that have emphasized the need to build modern security risk assessment frameworks for healthcare in Industry 5.0 \cite{Baz2023-zz}. During the COVID-19 pandemic, biotechnology sectors like healthcare, pharmaceutical companies, and government departments for public safety faced increased security breaches and cyber attacks \cite{He2021-tl} which emphasized the need to implement robust access control. Employing more robotics and automation in manufacturing in Industry 5.0 can also lead to vulnerabilities like data theft and security breaches which need to be addressed \cite{10200319}. With the increased amount of data sharing in the advanced solutions developed in Industry 5.0 using AI, there is an increased risk to the data \cite{Chander2022-sc} and hence developing frameworks to assess and manage these risks is crucial to the success of Industry 5.0. 

Besides the risks and vulnerabilities in the digital landscape, the affects of AI and technology also pose risk in the physical system. Supply chain and manufacturing sectors of the biotechnology Industry 5.0 deal with production, manufacturing as well as waste disposal in the biotech industry. Biotechnology Industry 5.0 sectors, mainly the pharmaceutical industry, is a major producer of heavy metals. With rapid advancement in applying technologies to supply chain management, the production and manufacturing of pharmaceuticals have skyrocketed. There is a need to assess the risk and take steps towards mitigating and managing the risks from heavy metal release into the environment \cite{Wang2022-us}. AI algorithms have showcased promising results on the detection of heavy metals like mercury, cobalt and chromium in blood and urine samples \cite{Yadav2023-dn}.  Machine learning has been used to develop security assessment and management frameworks for digital consumers and the energy industry \cite{Sharma2023-bv, Paltrinieri2019-xy} which emphasizes the role of ML and AI in developing secure and robust risk analysis systems in Industry 5.0. 

In biotechnology Industry 5.0, one of the most vulnerable components is data. Regulating the data access and implementing access controls on this crucial data is of utmost importance. Electronic record compliance in FDA regulatory frameworks emphasizes implementing access control in enterprises and industries \cite{noauthor_2023-ql}. The pharmaceutical sector of the biotechnology industry has been threatened and faced cyber attacks on multiple occasions \cite{noauthor_undated-mx}. There have been multiple data breaches that have put the privacy of people as well as the security of the population at high risk. It is, therefore, important to build access-controlled secure systems in Industry 5.0.

\subsection{Building Robust and Resilient Systems}
\label{resilent_subsec}

The biotechnology industry has witnessed a substantial increase in technology dependency over the decades. The role of modern technologies, especially AI, has been growing exponentially in all biotechnology sectors \cite{Holzinger2023-hs}. Although ML combined with IoT has led to some groundbreaking developments in biotechnology sectors like biometrics, healthcare, and agriculture, building resilient systems and adopting powerful ML techniques is an ongoing challenge. One of the major issues in building resilient and robust systems in the biotechnology Industry 5.0 is deploying the machine learning pipelines securely. This involves considering any retraining of the models that might be required after releasing into the production lines \cite{Heymann2022-qe}. Since machine learning models might need to be retrained on data continuously over time, some studies have proposed robust frameworks for continuous deployments of these ML models \cite{Jelassi2024-fs}. Biotechnology Industry 5.0 products that integrate IoT and AI models depend on these robust frameworks for commercialization. 

While protecting ML models is crucial, biotechnology Industry 5.0 can also exploit the power of these models in developing robust and resilient systems. Machine learning and deep learning techniques have been used to develop efficient frameworks for anomaly and outlier detection \cite{7344872, Hodge2004-eb,9065280}. Since the advent of the manufacturing industry, having humans in the loop has enhanced the reliability as well as efficiency of the products. Biotechnology Industry 5.0 deals with real-world problems that are crucial and high-risk. One example is during COVID-19 when production lines of vaccines were threatened \cite{Sanger2020-nv}. 

The ML models employed in agriculture, healthcare, pharmaceuticals, and supply chain sectors of the biotechnology Industry 5.0 are also at a very high risk for attacks. There have been studies that showcase the effectiveness of human-in-loop for human-machine teaming (HMT) in the healthcare sector. The study has effectively simulated the human-in-loop intelligent system in two case scenarios - 1. emergency care provider using voice-based synthetic assistants for 
MARCHE protocol scenario (massive hemorrhage) where it provides a step-wise treatment plan with no adverse effect to the patient and 2. Medic treating a patient follows the scenario in case 1 but involves extra steps of history taking and assessing the patient's airways where the caretaker takes less time to understand the instructions and thus fastens the treatment plan. This study has showcased how human-in-the-loop (HIL) can be used to enhance the performance of HMT systems \cite{Damacharla2020-ua}. Another study has quantified the effectiveness of HMTs and attempted to provide a benchmark to enhance human-machine collaboration \cite{8404030}. Keeping technology updated and current is another important aspect of building resilient systems to ensure smooth operations. AI/ML applications have been growing exponentially over the last decade. Updating models and replacing these models with better and recent models is crucial to developing systems that can prevail in the industrial revolutions. Voice-based assistants are now based on Large Language Models (LLMs) as opposed to Long Short Term Memory (LSTM) models which were quite heavily employed until the introduction of the transformer architecture \cite{Chen2023-tl}. These recently developed voice assistants are robust, reliable, and efficient and have a promising future in the manufacturing, production, as well as advertising sectors of Industry 5.0 \cite{November_undated-eq}. Some of the guidelines and practices discussed here for the development of robust and resilient systems are crucial for the integration of ML in product development and enhancing the security of current products in biotechnology Industry 5.0. 

\subsection{Monitoring and Incident Resolution}

MLSecOps in biotechnology Industry 5.0 ensures privacy and safety in products or services integrating machine learning with other technologies like IoT and cloud computing. Developing and defining standards to use modern technologies ensures standardization, global usage, and effective commercialization. So far in this section, we have discussed risk assessment and management of ML pipelines as well as practices that pave the way for the development of robust and resilient systems in Industry 5.0. However, ML models are sensitive and the results can easily fluctuate on outlier data points \cite{Sharma2023-az}. The vulnerabilities of ML models make it crucial to include monitoring the models in the practices of MLSecOps in Industry 5.0. This chapter has discussed how the biotechnology Industry 5.0 deals with crucial data that affects the economy, industry, as well as end users. One of the ways to make sure that the ML models are resilient is to probe these models for weaknesses. Adversarial Machine Learning (AML) is among the most effective techniques for checking ML pipelines for attacks, threats, and vulnerabilities. AML has been used to test the defense mechanism of ML techniques in IoT environments \cite{Alkadi2023-xm} and intrusion detection systems \cite{Alatwi2021-tr,9454214}. 

While probing the ML models is an effective technique to find weaknesses of the developed systems in Industry 5.0, the models also need continuous training and deployment as discussed in section \ref{resilent_subsec}. It is therefore important that the models be monitored. Continuous monitoring of the ML integrated system not only helps protect them against attacks but also helps identify any drift or abnormalities in functioning. Machine learning has promising growth in the supply chain management sector of biotechnology Industry 5.0. Frameworks for continuous monitoring in the supply chain post-deployment have been proposed \cite{Eck2022-ar}. In biotechnology Industry 5.0, we not only need to monitor ML models but also the data. Any data alterations can lead to abnormal behavior of ML models in an ecosystem. Therefore, monitoring the data to prevent any data poisoning is also crucial in Industry 5.0, especially the bioinformatics and genetic sectors of biotechnology Industry 5.0. Access control over who can retrain and deploy models, checking the updated model for performance drifts, and control over who can get access to data are some of the techniques to handle and detect data poisoning \cite{Constantin2021-xu}. Monitoring ML ecosystems and implementing practices to deal with threats or attacks ensures a robust, enduring, and safe biotechnology Industry 5.0. 

This section has discussed effective MLSecOps practices that can be employed to advance the biotechnology Industry 5.0 while ensuring privacy and security. The section has also given an overview of the importance of monitoring ML systems as well as the data used in AI/ML environments. Access control and monitoring practices concerning ML integration ensure that the biotechnology Industry 5.0 has a positive impact on the economy, strong research, and end-user acceptance of innovative products. Towards the end of this chapter, the focus is drawn toward ethical and social responsibilities in Industry 5.0 and providing key points to assist decision-makers in developing compliance and regulatory frameworks for biotechnology Industry 5.0. 

\section{Ethical Considerations and Social Responsibility}\label{sec7}

Biotechnology Industry 5.0 focuses on developing human-centric products. While the latest technological advances like machine learning, natural language processing models, and computer vision models are being integrated, Industry 5.0 has a goal of personalization. Personalizing medicine, workouts, meals, etc. is advancing in all sectors of Industry 5.0. This chapter has outlined how Industry 5.0 will witness the growth of integrating ML with other technologies in all sectors. This will lead to biotechnology Industry 5.0 contributing more to the global economy. However, section \ref{threat_landscape} has provided an analysis of the threats that are current in the biotechnology Industry 5.0. Although, Industry 5.0 has laid out some regulations discussed in section \ref{sec5} the biotechnology Industry 5.0 is still vulnerable. In this section, we will highlight the importance of considering social responsibilities and ethical considerations in biotechnology Industry 5.0. 

\subsection{Ethical Use of Biotech Data: Balancing innovation with privacy rights}

The biotechnology industry has a plethora of data. Biometrics, genetic data, patient notes, and medical device data are some of the most common examples of data in the biotechnology industry. This industry is expected to grow by 13.96\% approximately \cite{noauthor_undated-ek}. Data is a huge source of income and Industry 5.0 aims to build products for the end users that harness the power of ML algorithms on this data. The data is crucial and needs to be secured while being employed in the product development stage. Section \ref{intro} has discussed how the data must be protected and secured while being used in ML algorithms as well as while being stored in the cloud. There is however one more aspect to this. The users must be aware that their data is being used. Many biotechnology companies monetize their data by selling it to other companies for product development \cite{Konieczka2023-ly}. This however raises a concern - \textbf{Is the user aware that their data is being sold?}. There have been multiple reports of private data being sold without the consent of users \cite{Fung2024-ef} and this puts the privacy of individuals at risk. However, there have been introductions and amendments to consumer data privacy acts globally that make sure that users are aware of where their data goes \cite{Nahra2023-ez, Nahra2023-uu, noauthor_2021-qs,noauthor_2021-zz}. The users should not only be aware of the fact that their data is being sold but there should be transparency of the data activities. The user should have full information on how the company is using the data internally and if the data is being provided to external sources \cite{Boue2018-kr}. This builds trust among users and enterprises as well as assures that the user is compliant with the data being used. To build a balanced ecosystem of innovation and security the biotechnology Industry 5.0 needs to ensure that they are following cybersecurity best practices. This minimizes the vulnerability of the ML models, software, and the data. Section \ref{cybersecuritty_regulations} has discussed the cybersecurity regulations and policies. These regulatory frameworks ensure consumer as well as industry protection and growth.

\subsection{AI Ethics in Biotechnology: Addressing potential biases and ethical dilemmas}

The scope of this chapter is to focus on biotechnology Industry 5.0. Personalization and developing technology that is human-centric has been the goal of Industry 5.0 and integrating ML with other technologies like cloud computing and IoT has made this goal achievable. However, ML algorithms are vulnerable which makes them an easy target for malicious activities. MLSecOps in biotechnology Industry 5.0 focuses on developing secure biotechnology practices by securing ML algorithms as well as enhancing security by employing ML models. This has been described in section \ref{sec3}. AI models perform based on the data they are trained on. The biases of these data get transferred to the models implicitly. The biometric sector of biotechnology Industry 5.0, for example, has developed advanced facial recognition software. However, there have been reports of racial bias in this software where they have failed to recognize people of color \cite{SITNFlash2020-nh}. Similarly, facial recognition systems also exhibit demographic biases \cite{9086771}. Gender bias in research and clinical trials has been reported where male-centric research is given more funding compared to female-centric research \cite{Chilet-Rosell2014-ab, Orbach2022-ab}. There have been attempts to mitigate these biases \cite{Huang2023-lb}. Addressing these issues in the biotechnology Industry 5.0 which includes the implicit bias in ML systems will result in a secure and privacy-oriented Industry 5.0. Although, identifying the biases and mitigating them in AI is an important step towards a more safe biotechnology Industry 5.0 it is crucial that regulatory frameworks be formed and the industry complies with these frameworks in their practices. Section \ref{cybersecuritty_regulations} has described the current as well as emerging cybersecurity frameworks. These frameworks ensure precautionary steps are taken and there is transparency between the end user and the companies regarding how their data is being used. Taking responsibility for one's actions is a step towards empathy and improvement. Biotechnology Industry 5.0, especially the healthcare sector, demands that not only precautionary measures be taken but in case of any errors, it is important that the industry takes responsibility and makes corrections of these errors. Liability frameworks that consider the immense integration of AI in healthcare and medicine ensure that there is a balance between innovation and safety \cite{Maliha2021-un}.

\subsection{Social Implications: The impact of biotech advancements on society.}

We are well aware that change incurs some impact. While we try to maximize the positive impact of any development, there are always some negative impacts associated with it. Similarly, with advanced technology being at the center of the biotechnology Industry 5.0, we are looking at labor force change, where reskilling and upskilling will be critical. With advancements in supply chain automation in Industry 5.0, human jobs are expected to decelerate \cite{Demir2019-cb}. This may be a point of concern for society where more and more automation can cut out the jobs of humans and their livelihoods. Biotechnology Industry 5.0 must focus on reskilling and upskilling the talented workforce, particularly in areas where jobs will be impacted by automation. Besides this, another social impact of Industry 5.0 can be the cost. With advanced medical devices in the healthcare industry as well as AI-enabled gene therapies, the affordability of the industry can decrease \cite{The2023-fm}. However, biotechnology Industry 5.0 focuses on employing cutting-edge technology with a great focus on sustainability \cite{Low2020-rb}. This not only makes the products from biotechnology Industry 5.0 like pharmaceuticals, healthcare, clinical trials, etc. affordable but also promotes the bioeconomy. 

Biotechnology Industry 5.0 possesses great potential to improve human life on many fronts. The healthcare industry has displayed great potential at integrating technology to improve patient care, remote patient monitoring, and using voice-based assistants in surgical processes as mentioned in section \ref{resilent_subsec}. Keeping societal well-being and ethical responsibilities in consideration, biotechnology Industry 5.0 will boom at providing human-centric solutions to current problems in all sectors. Understanding the vulnerabilities and the role of MLSecOps in the biotechnology Industry 5.0, we have provided a perspective of the ethical and social responsibilities in this section that must be adhered to develop sustainable, robust, and resilient systems.

% \section{Policy Recommendations for Decision Makers}\label{sec8}
% \titus{We may not need this section, so let's see how much space we have left when we get here.}

\begin{backmatter}
\section{Conclusions}
The integration of MLSecOps into the biotechnology Industry 5.0 is not just beneficial but imperative for safeguarding against increasingly sophisticated cyber threats. This chapter has underscored the critical vulnerabilities within the industry and presented a case for adopting MLSecOps practices. By embracing a proactive approach to cybersecurity, leveraging advanced ML techniques, and adhering to regulatory and ethical standards, the biotechnology sector can fortify its defenses and ensure the privacy and security of its data and systems. Looking forward, the industry must continue to evolve its security practices in tandem with technological advancements, ensuring a resilient and robust defense mechanism that can adapt to new challenges. In doing so, the biotechnology Industry 5.0 can achieve its full potential, driving innovation while safeguarding the invaluable data at its core.

\section*{Abbreviations}

\begin{abbrvlist}[DMEM-FBS]
\item[ML] Machine Learning
\item[IoT] Internet of Things
\item[AI] Artificial Intelligence
	
\end{abbrvlist}

% \section*{Appendix A}

\begin{authordetails}
	
	% Author details will always appear the end of the chapter in the final version of the chapter
	
	\author{Naseela Pervez$^{1}$ and Alexander J. Titus $^{1, 2, 3}$}
        \address[1]{In Vivo Group, Los Angeles, USA}
        \address[2]{Information Sciences Institute, University of Southern California, Los Angeles, USA}
        \address[3]{Iovine and Young Academy, University of    Southern California, Los Angeles, USA}
	\address{*Address all correspondence to: publications@theinvivogroup.com}
	%	
	% \IntechOpentext{\textcopyright\ \the\year{} The Author(s). License IntechOpen. This chapter is distributed under the terms of the Creative Commons Attribution License (http://creativecommons. org/licenses/by/3.0), which permits unrestricted use, distribution, and reproduction in any medium, provided the original work is properly cited.}
	
% 	% Note: The copyright year will be changed accordingly during production to correspond with the year of publication.
	
\end{authordetails}

\bibliographystyle{vancouver}
\bibliography{bibliographies}

\end{backmatter}

\end{document}